\documentclass[titlepage,reprint,superscriptaddress,prl]{revtex4-1}
\usepackage{amsmath}
\usepackage{amssymb}
\usepackage{graphicx}
\usepackage{color}

\newcommand{\avg}[1]{\langle #1 \rangle}
\newcommand*{\GtrSim}{\gtrsim}

\newcommand{\Ki}{\mathrm{K}^+}
\newcommand{\Cli}{\mathrm{Cl}^-}		
\newcommand{\K}{\mathrm{K}}
\newcommand{\Cl}{\mathrm{Cl}}

\def\fnum@figure{\figurename\nobreakspace\textbf{\thefigure}}

\begin{document}

\title{Dehydration as a Universal Mechanism for Ion Selectivity in Graphene and Other Atomically Thin Pores}

\author{Subin Sahu}
\affiliation{Center for Nanoscale Science and Technology, National Institute of Standards and Technology, Gaithersburg, MD 20899}
\affiliation{Maryland Nanocenter, University of Maryland, College Park, MD 20742}
\affiliation{Department of Physics, Oregon State University, Corvallis, OR 97331}

\author{Massimiliano Di Ventra}
\affiliation{Department of Physics, University of California, San Diego, CA 92093}

\author{Michael Zwolak}
\affiliation{Center for Nanoscale Science and Technology, National Institute of Standards and Technology, Gaithersburg, MD 20899}
\email{mpz@nist.gov}

\begin{abstract}
Ion channels play a key role in regulating cell behavior and in electrical signaling. In these settings, polar and charged functional groups -- as well as protein response -- compensate for dehydration in an ion-dependent way, giving rise to the ion selective transport critical to the operation of cells. Dehydration, though, yields ion-dependent free-energy barriers and thus is predicted to give rise to selectivity by itself.  However, these barriers are typically so large that they will suppress the ion currents to undetectable levels. Here, we establish that graphene displays a measurable dehydration-only mechanism for selectivity of $\Ki$ over $\Cli$. This fundamental mechanism -- one that depends only on the geometry and hydration -- is the starting point for selectivity for all channels and pores. Moreover, while we study selectivity of $\Ki$ over $\Cli$, we find that dehydration-based selectivity functions for all ions, i.e., cation over cation selectivity (e.g., $\Ki$ over $\mathrm{Na}^+$). Its likely detection in graphene pores resolves conflicting experimental results, as well as presents a new paradigm for characterizing the operation of ion channels and engineering molecular/ionic selectivity in filtration and other applications.
\end{abstract}

\maketitle

Ionic transport through protein pores underlies many biological processes.
Various factors, such as the presence of charges and dipole moments,
structural transitions of the pore, dehydration of ions, make
the dynamics of biological pores very complex {\cite{Doyle98-1,noskov2004,noskov2007,Rasband2010,varma2008,eisenberg1999structure,varma2011}. Solid-state nanopores can delineate these contributions to transport.
For instance, a competition of hydrated ion size and pore radius was predicted to result 
in a drastic drop in ionic conduction \cite{Zwolak09-1,Zwolak10-1} -- in a way analogous to quantized conductance in electronics -- and subsequently observed in graphene laminates~\cite{Joshi2014}. These studies \cite{Zwolak09-1,Zwolak10-1}, as well as others \cite{sint2008,song2009,richards2012,richards2012quantifying}, show that, as expected, ions have different free energy barriers due to dehydration, allowing dehydration to yield selectivity on its own. However, the large free-energy barriers typical of ions entering long, narrow pores greatly suppress the current from all ions, making this selectivity difficult to resolve \cite{Gouaux2005,sahu2017}.

\begin{figure}[h!]
\includegraphics{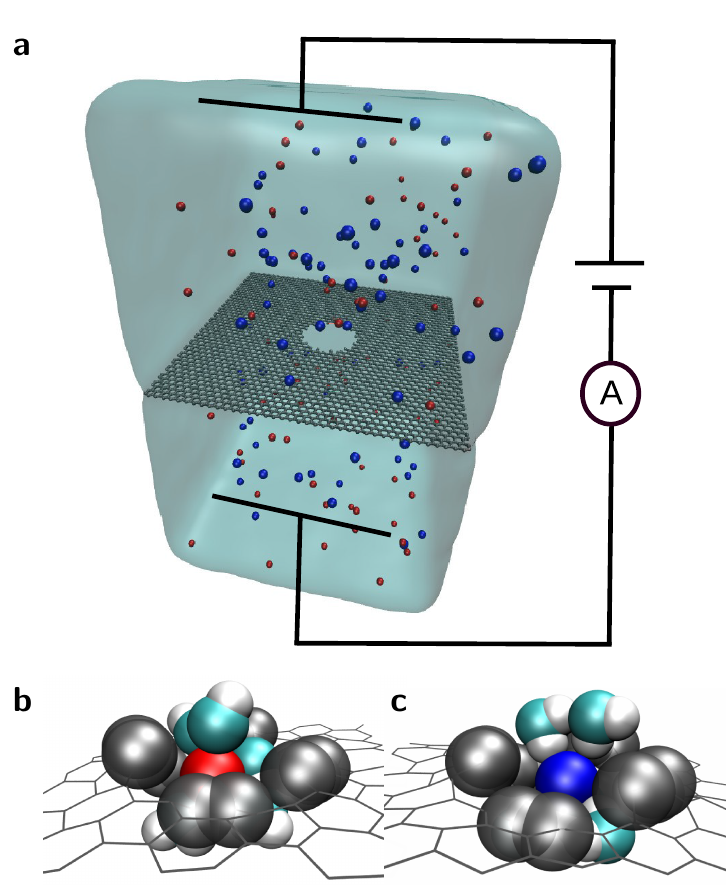}
\caption{\label{fig:sim} {Schematic of ion transport through a graphene nanopore.} (a) An applied bias across the membrane drives an ionic current so that, when the pore is sufficiently small, the ions have to partially shed their hydration layer in order to translocate through the pore. Red and blue spheres represent $\Ki$ and $\Cli$ ions,  respectively. Water is shown as a continuum even though we use an all-atom model for this work. (b) A potassium ion and (c) a chloride ion crossing the pore maintain significant hydration, unlike in long, narrow pores. Carbon at the edge of the pore, oxygen, and hydrogen are gray, cyan, and white spheres, respectively.}
\end{figure}

Due to its atomic thickness, one expects graphene may provide a suitable membrane to detect and exploit dehydration-only selectivity, as we will discuss here. Several recent experiments give tantalizing results in this direction, where, for instance, Refs.~\onlinecite{garaj2010} and~\onlinecite{Jain2015} suggest they may have seen dehydration-based selectivity. For instance, Ref.~\onlinecite{garaj2010} finds that the graphene membrane they use has a small -- but measurable -- conductance when no pores are present, which they attribute to defect channels. Salts, XCl, with X=Cs$^+$, Rb$^+$, $\Ki$, Na$^+$ and Li$^+$, give conductances that slightly deviate from what their bulk conductivities would predict. However, as their Table 1 and our Table S3 in the Supporting Information (SI) show, the discrepancy is small (less than 50~\%). In fact, if the defect channels -- the structures of which are not known -- are cation selective due to the presence of negative charges, the {\em bulk properties} almost perfectly predict the observations. Moreover, the difference in hydration energies is large and not expected {\em a priori} to give such a small deviation in conductance (see the SI). These considerations suggest that dehydration is likely {\it not} responsible for the small deviation in membrane-only conductance in Ref.~\onlinecite{garaj2010}.

In a similar vein, Ref.~\onlinecite{Jain2015} finds weak monovalent cation selectivity over divalent cations for sub-2 nm pores. However, the pores had a large variability in conductance over time (see their Fig. 3). Within these uncertainties, the data may be explained by the different mobilities alone, as we show in Table S4 in the SI. Moreover, Ref.~\onlinecite{rollings2016} recently found highly selective ($\Ki$ over  $\Cli$) graphene pores, where a charged graphene membrane is the likely cause since the selectivity persists for large pore sizes. These membranes also showed a weak monovalent cation selectivity over divalent cations for pores that were 2 nm to 15 nm in diameter. Indeed, for pores in the 2 nm to 6 nm range, the selectivity was of the same magnitude as that observed by Ref.~\onlinecite{Jain2015}, where it is also indicated that their pores are charged. These considerations suggest, as well, that dehydration is not the cause of selectivity in Ref.~\onlinecite{Jain2015}.

The strongest evidence for dehydration-only selectivity in transport comes not from nanopores but from nanochannels in graphene laminates. Ref.~\onlinecite{Joshi2014} shows that channels with a height of $\approx 0.9$ nm allow atomic and small molecular ions to permeate but exclude larger molecular ions, and this exclusion is correlated with hydrated radius (as those authors discuss, adsorption may be playing a role in addition to steric hindrance of the water shell). This height, though, is above the scale necessary to mimic and understand biological ion channels. 

Another recent study, however, finds selectivity of $\Ki$ over  $\Cli$ in porous graphene \cite{OHern2014}. This selectivity drops rapidly when the (mean) pore diameter increases by about 0.1 nm to 0.2 nm. This was attributed to the presence of charges. However, such a sharp feature is indicative of an atomic scale process, of which dehydration is the obvious culprit. Moreover, this is selectivity of a mono-atomic ion over another (with almost identical mobilities), just as in biological systems. We will show that the selectivity they find ($\approx 1.3$ for a distribution of pore sizes, giving $\approx 1.8$ to $2.5$ for a pore radius of $\approx 0.2$ nm, see our SI) is consistent with the partial dehydration ions experience when going into single atom thick pores with a radius in the sub-nanometer scale. We believe that this is the best evidence of dehydration-only selectivity in nanopore experiments so far. \\

{\noindent \bf Results and Discussion}\\
Atomically thin pores -- pores that are one to a few atoms thick, such as graphene, MoS$_{2}$, or hexagonal boron nitride -- display more intricate behavior when the pore radius reaches the scale of the hydration layers compared to long pores and channels. 
Molecular dynamics
(MD) simulations show that the ionic current through graphene
pores (see the schematic in Fig.~\ref{fig:sim}) exhibits nonlinear behavior as the pore radius, $r_{p}$, is
reduced to the sub-nanometer scale. This nonlinear behavior is seen
as a rapid drop in the ionic current, Fig.~\ref{I-ratio}(a,b), and an excess noise in the current, 
see the SI. At these length scales the pore
edge begins to deform the hydration layers, increasing
the energy barrier for ions to cross the pore.

\begin{figure}[t]
\includegraphics{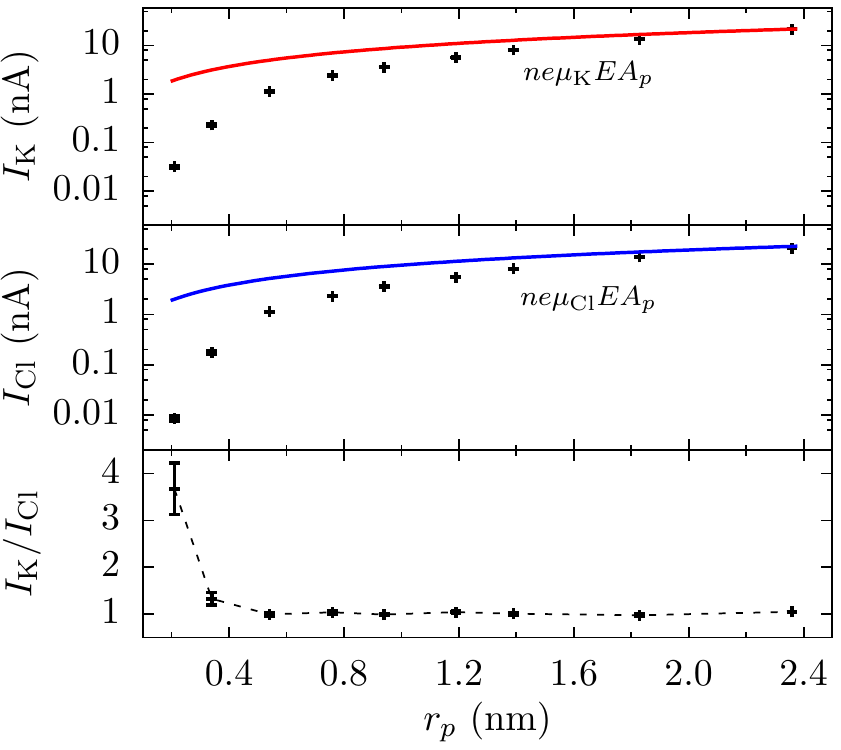}\label{KCL}
\caption{\label{I-ratio} Selective behavior of graphene pores as a ratio of cation ($I_\K$) to anion ($I_\Cl$) currents versus the pore radius (as determined by MD simulations). For large pores ($r_p \GtrSim  2$ nm), the ionic current is due to bulk flow but limited by the pore's cross-sectional area, $A_p$, available for transport. In this situation, the contribution to the current from species $\nu$ is proportional to $e z_\nu n_\nu \mu_\nu$ (times $A_p$ and the electric field, $E$, in the vicinity of the pore due to a 1 V applied bias, see Fig. S3), where $z_\nu$ is its valency, $n_\nu$ its number density, and $\mu_\nu$ its mobility (shown as red and blue lines for $\Ki$ and $\Cli$, respectively). Thus, for potassium and chloride, the ratio is $I_\K/I_\Cl = \mu_\K/\mu_\Cl \approx 1$, as the relative mobility of $\Ki$ to $\Cli$ is 1/1.04 . When the pore is very small, though, the relative cationic contribution to the total current increases. This is due to a higher dehydration energy of the anion compared to the cation. Connecting lines are shown as a guide to the eye. Error bars are $\pm 1$ Block Standard Error (BSE) everywhere unless indicated otherwise (details are discussed in the SI).}
\end{figure}

 \begin{figure}[h!]
 \centering
 \includegraphics{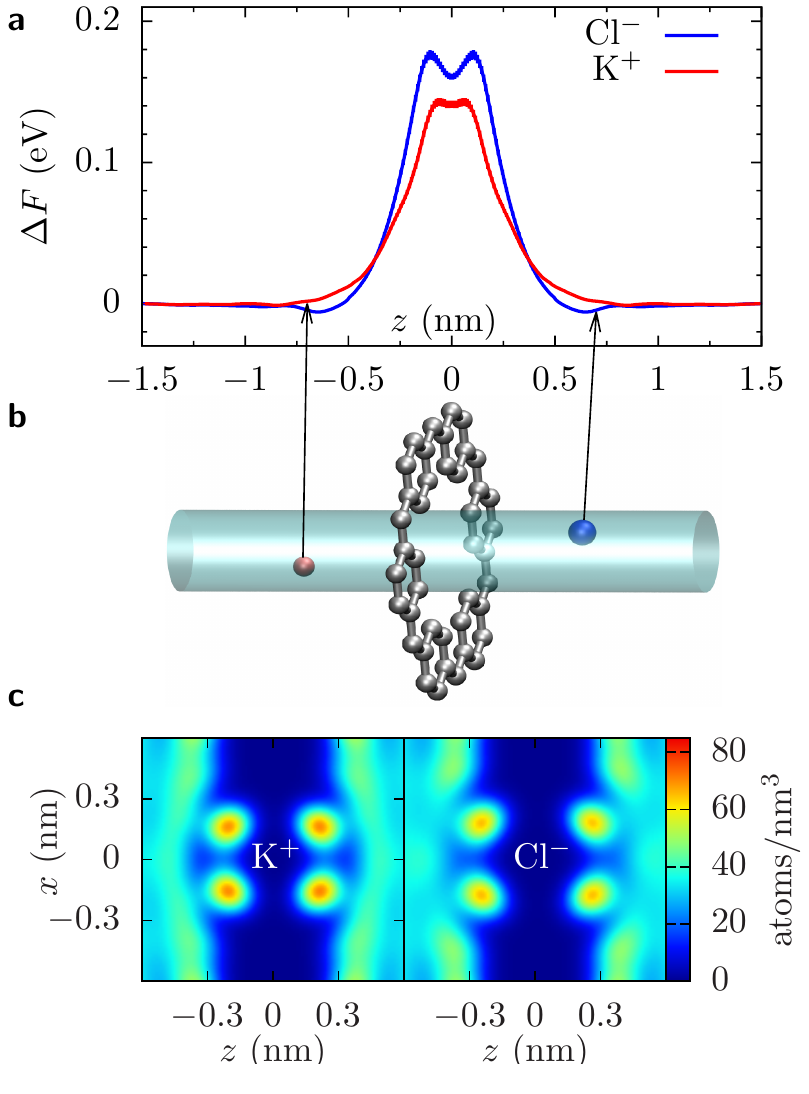}
  \caption{\label{ABF} 
 Free energy and hydration in graphene pores. (a) The free energy change of an ion to go from bulk into a pore with radius $r_p=0.21$ nm versus the ions' $z$ coordinate. (b) Schematic of the simulation showing the cylindrical region where a $\Ki$ (red sphere) or a $\Cli$ (blue sphere) ion is confined for the ABF calculation. In the smallest pore and at 1 V, the relation $I_\K/I_\Cl=A_0 e^{(\Delta F_\K -\Delta F_\Cl)/k_B T} \approx 3.5$, where $A_0 \approx  1$, $k_B$ is Boltzmann's constant, and $T$ is temperature, implies that the free energy change of $\Cli$ should be about 30 meV larger than that for $\Ki$. This is consistent with the ABF calculation (between 20-40 meV depending on where we measure the difference).
(c) Oxygen density from water around a potassium or a chloride ion fixed at the center of the $r_p=0.21$ nm pore. Within the first hydration layer, there are $\avg{n}=5.2$ and $\avg{n}=5.8$ waters around $\K^+$ over $\Cl^-$, respectively, which is approximately $1/4$ less than in bulk in both cases. 
See the SI for more data regarding $\Delta F$ and the density plots. Error bars in (a) are $\pm 1$ BSE.}
\end{figure}

In particular, graphene pores with $r_p=0.21$ nm and in 1 mol/L KCl display {selectivity of $\Ki$ over $\Cli$ despite containing no charges or dipoles, as shown by the MD simulations in Fig. \ref{I-ratio}(c)}. That is, no electrostatic repulsion or specific interaction with the membrane is needed for its appearance. The selectivity and sharp rise in resistance was also seen for lower concentrations of KCl (see the SI), i.e.,  this behavior is not due to ion-ion interactions or some other many-body effect.

To quantify the energetic differences of $\Ki$ and $\Cli$, we perform Adaptive Biasing Force (ABF) calculations for the smallest pore size. The ABF results, Fig. \ref{ABF}(a,b), demonstrate that -- without an applied voltage -- the $\Ki$ is between 20-40 meV (depending on where we measure the difference) more favored to be in the pore compared to $\Cli$, although both have to pay a substantial energy penalty. These equilibrium energy barriers are determined by the partial dehydration of the ions. When ions are in the atomically thick pore, water molecules can rearrange just outside of the pore to maintain significant hydration of the ion (see Fig. \ref{ABF}c). Ions only have to lose about $1/4$ of their inner hydration water molecules in order to be at the pore center. When the pore radius is smaller than the radius of the first hydration layer, a much larger dehydration occurs (about $3/4$) in long pores, as water molecules can only be at opposite ends of the ion (see Refs.~\onlinecite{Zwolak09-1,Zwolak10-1}). This will lead to related, but different, nonlinearities in the current.

\begin{figure}[h!]
\includegraphics{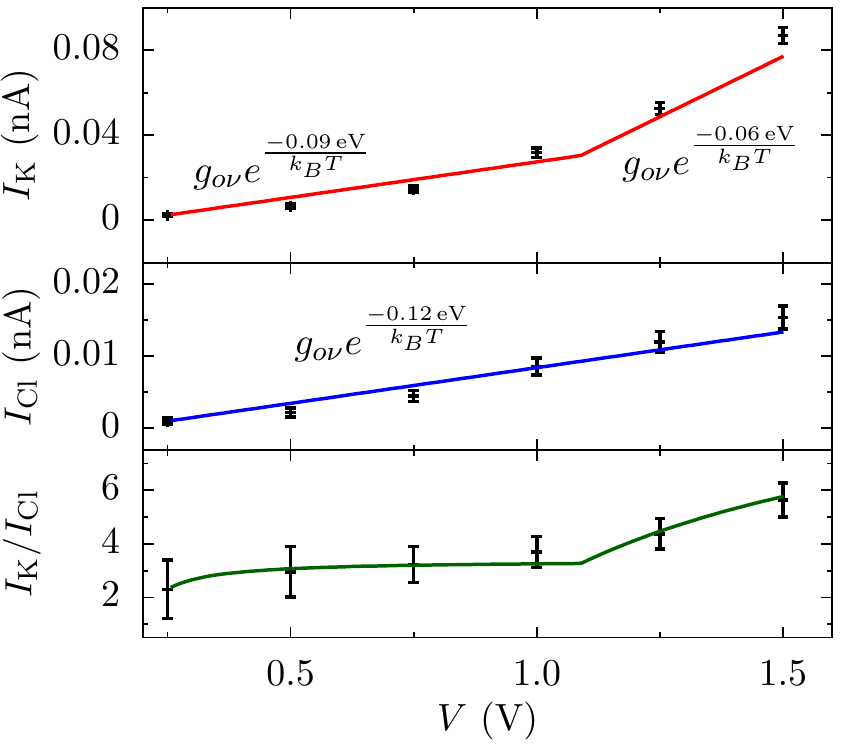}
\caption{\label{IV} Current and selectivity versus voltage. Current-voltage characteristics for $\K^+$ and $\Cl^-$ for the pore of radius 0.21 nm at 1 mol/L KCl (top panels). The  selectivity of $\K^+$ over $\Cl^-$ in the same pore (bottom panel). For small voltages, the ions display roughly linear behavior. Thus, we fit a piece-wise linear model for current taking each region to be linearly related to the differential conductance $g_\nu=e z_\nu n_\nu e^{(-\Delta F_\nu/k_B T)} \mu_\nu A_p/L = g_{0\nu} e^{(-\Delta F_\nu/k_B T)}$. This piece-wise local conductance model (solid lines) can not capture all the features observable in the MD simulations (data points), but it captures the suppression of the current by the free energy barrier and how that barrier changes with voltage. Around $(1.1\pm0.1)$ V the energy barrier for potassium drops from $\approx (0.09\pm 0.004)$ eV to $\approx (0.06\pm0.002)$ eV, whereas that of chloride remains approximately constant at $\approx (0.12\pm0.001)$ eV. This results in sharper rise in the potassium current and, hence, selectivity. Error bars are $\pm 1$ BSE.}
\end{figure}

The difference in energy barriers is due to a smaller dehydration energy of $\Ki$ compared to $\Cli$, which can be estimated from the relation, 
\begin{equation}
\Delta E_\nu(r_p) = f_{1\nu}(r_p)E_{1\nu} \label{eq:fE},
\end{equation} 
where $f_{1\nu}$ is the fractional dehydration in the first hydration layer and $E_{1\nu}$ is energy of the first hydration layer \cite{Zwolak09-1,Zwolak10-1}. This energy penalty is about 0.27 eV for $\Ki$ and  0.37 eV for $\Cli$ when using $f_{1\nu} \approx 0.25$ (see Fig.~\ref{ABF}(c)). Moreover, the fraction $f_{1\nu}$ can be approximated purely on geometric grounds: When an ion is in the center of the smallest pore, water molecules can not enter the pore due van der Waals (vdW) interactions with the carbon and the ion. The area available for water is composed of two spherical caps of radius $r_{1\nu}$ -- the radius of the first hydration layer -- on either side of the membrane. The fractional area encroached by membrane is $d/2 r_{1\nu}$ where $d$ is the thickness of the membrane as seen by the water molecules near the pore edge. This thickness is $d \approx 0.2$ nm, slightly smaller than the distance between water and the graphene membrane, as water molecules can move into the corrugated openings at the pore edge. This gives  $f_{1\nu} \approx 0.3$ for both ions, when using $r_{1\nu} = 0.33$ nm and $0.31$ nm for $\Ki$ and  $\Cli$, respectively \cite{Zwolak10-1}, in agreement with the full MD calculation. 

The free-energy difference from the ABF calculation for each ion is about half of the value estimated using Eq.~\eqref{eq:fE}. To show why, we examine the net water dipole in the first layer when the ion is inside and outside the pore. The net dipole is reduced by about 5~\% to 10~\%. {Therefore, when about 25~\% of the water is lost, the remaining waters orient more strongly and the energy barrier is not 25~\% of the first hydration layer energy, but less than half of that.} The energy barrier is thus qualitatively in agreement with what Eq.~\eqref{eq:fE} predicts but reduced due to stronger orientation of the remaining water dipoles (which is possible because interference from water-water interactions decreases). Finally, we note that other effects, such as polarization of graphene and different functional groups and/or pore shape (i.e., different atomic scale structure and graphene edge type), will change the observed barrier. However, when charges or static dipoles are absent, the main contribution to the free energy barrier is due to fractional dehydration -- as qualitatively captured by Eq.~\eqref{eq:fE}. These barriers will differ from ion to ion, giving rise to weak selectivity (depending on the ion types, this will be cation over anion, anion over cation, cation over cation, etc.). This selectivity will thus appear in other atomically thin pores (pores that have a thickness of one to a few atoms), such as hexagonal boron nitride and  MoS$_{2}$.

The selectivity also depends on the applied voltage, as seen in Fig. \ref{IV} (see also the SI). The increasing selectivity from 0.25 V to 1 V is due to the lower barrier for $\Ki$ versus $\Cli$. In this region, the current contribution from each ion species $\nu$ can be fit with the form
\begin{equation}
I_\nu = g_\nu V + c_\nu \label{eq:Curr},
\end{equation}
where
\begin{equation}
g_\nu  = e z_\nu n_\nu e^{(-\Delta F_\nu/k_B T)} \mu_\nu A_p/L \label{eq:cond}
\end{equation}
and $c_\nu$ is a constant. This form of the current cannot capture all of the intricate behavior, but it does capture the main feature -- the suppression of the current by the dehydration free energy barrier. The magnitude of $c_\nu$ reflects behavior below 0.25 V, which is a regime that is difficult to reach with all-atom simulations. Given $c_\nu$, the differential conductance, Eq. \eqref{eq:cond}, determines the selectivity as the voltage increases. This conductance depends on the valency, density, mobility, and free-energy barrier of ion $\nu$, as well as the electric field $E$ and pore area $A_p$. Taking the electric field to be $V/L$ with $L=1$ nm (in agreement with the fields found from the MD simulations) and $A_p=\pi r_p^2$, the nonequilibrium free-energy barriers are ($0.09\pm 0.004$) eV and ($0.12\pm 0.001$) eV for $\K^+$ and $\Cl^-$, respectively. These values are slightly less than the equilibrium free-energy barriers from the ABF calculation in Fig. \ref{ABF}. As we discuss below, this is likely because of water polarization in the pore that helps ions to move through. The value of other parameters, especially $A_p$ and the voltage range, also affect the extracted free-energy barriers both in equilibrium and out of equilibrium (computationally, for instance, sampling a smaller area in Fig.~\ref{ABF} (a,b) will lower the barrier, and fitting the I-V curve at lower voltages will increase the barrier extracted). We note, though, that the difference in free-energies between $\Ki$ and $\Cli$ is the same in equilibrium and at low voltages. 

Using the parameters for $\K^+$ and $\Cl^-$ in Eq. \eqref{eq:Curr}, the selectivity will increase with voltage until it saturates at 
\begin{equation}
\frac{I_\K}{I_\Cl} = \frac{g_\K}{g_\Cl},
\end{equation}
which fits well with the selectivity data from MD, as shown in Fig. \ref{IV}. However, we see that at $\approx 1$ V, the selectivity starts to increase still further. Indeed, the differential conductance of $\K^+$ increases at $\approx 1$ V due to an effective lowering of the barrier to ($0.06\pm 0.002$) eV. This lower barrier is due to a polarization-induced chaperoning of the ion across the pore. As the local electric field increases with voltage, water becomes increasingly polarized in the vicinity of the pore and also increases its density there. When, for instance, $\Ki$ sees a pore containing polarized water where the oxygen atoms are pointing towards the ion, it can more easily move to the pore edge and be taken through to the other side. The $\Ki$ responds to this at lower voltages than $\Cl^-$ due to -- surprisingly -- its vdW interaction with carbon, which allows the charging layer of $\K^+$ to be closer to the membrane than $\Cl^-$ (see Fig.~S-3(a), which shows the voltage drop is mostly on the $\Cl^-$ side).

%\section*{Discussion}
There are thus two facets of selectivity: the mechanism and the nonlinear features with voltage. The former depends on the dehydration energy and the pore geometry, and the latter depends on structural changes of water in the pore as the voltage increases (as well as vdW interactions). Since both are dependent only on geometry and ion characteristics in bulk water, these features will be present for other atomically thin pores and channels. This mechanism for nonlinearities and selectivity gives  simple predictions that can be tested experimentally by examining multilayer graphene~\cite{sahu2017}, MoS$_2$ pores, or other pores.

%----------------------------------------------------------------------------------------------
%--------------------------------------------------------------------------------

Moreover, selectivity has recently been observed in graphene membranes \cite{OHern2014,rollings2016}. In Ref.~\onlinecite{OHern2014}, the selectivity is weak, giving an average ratio of translocation rates of about $1.3 \pm 0.1$ for potassium to chloride at zero applied bias but in the presence of a concentration gradient across the membrane (see the SI). This experiment has a distribution of pore sizes for each membrane. Assuming the smallest pores make a negligible contribution to the current -- and thus do not affect the average selectivity even though they are themselves selective -- and the largest pores have no selectivity, the experimental observations yield a selectivity in the range $\approx 1.8$ to $2.5$ for a $r_p=0.21$ nm pore. This is larger than potential variation due to effective mobilities and agrees well with the dehydration-induced selectivity of $\approx 2.3$ that we find at low voltages. In addition, using the free energy barriers and areas from MD, we can compute the expected selectivity for the experiment. It comes out to be $\approx 1.5$, also in excellent agreement with observations (see the SI). In Ref.~\onlinecite{rollings2016} the selectivity is much higher ($\approx 100$), even for large pores ($r_p$ in the 1 nm to 10 nm range, i.e., well above the size of hydration layers). In both experiments, though, the selectivity was attributed to the presence of charges on the graphene.

Our results indicate, however, that weak selectivity is the consequence of different dehydration energies and fractional removal of water (determined by the hydrated ion size and pore geometry) -- i.e., not due to local charges or dipoles, which would result in the much higher selectivities observed in Ref.~\onlinecite{rollings2016} and in biological pores.  While charges/dipoles can not be ruled out as an explanation for the results of Ref.~\onlinecite{OHern2014}, they would need to be far from the pore, really small in magnitude, or only present in a minority of pores in order to generate weak selectivity. That is, charge selectivity is much stronger in general and also less sensitive to pore radius. This would leave unexplained why a small increase in pore radius eliminates selectivity. Screening can play a role here, suggesting that changing the molarity of the solution -- thereby changing when selectivity disappears as the pore radius increases -- is a simple test to determine the presence of local charges/dipoles.

The other effects we predict -- nonlinear features in the current-voltage characteristics and enhanced selectivity -- can be observable experimentally if the structural and electronic integrity of the membrane can be maintained at  higher voltages (e.g., biases in excess of about 250 mV start to degrade some graphene membranes \cite{garaj2010}. Polarization-induced chaperoning, though, should be present even in other thin membranes that may keep their integrity to higher voltages). These results elucidate selectivity and dehydration effects in transport when going from the nanoscale \cite{garaj2010,merchant2010,Schneider2010,sathe2011,wells2012} to sub-nanoscale channels and pores. It should help design and understand the behavior of solid-state pores that serve as hosts for other nanoscale probes, i.e., localizing and interrogating molecules or nanoscale structures, such as DNA sequencing with ionic \cite{Kasianowicz1996-1,Akeson1999-1,Deamer2000-1,Vercoutere2001-1,Deamer2002-1,Vercoutere2002-1,Vercoutere2003-1,Winters-Hilt2003-1,Vercoutere2002-1}
or transverse electronic currents \cite{Zwolak05-1,Lagerqvist06-1,Lagerqvist07-2,Zwolak08-1,Krems09-1,Tsutsui10-1,Chang10-1}.

Moreover, selectivity in biological ion channels is complex and due to many competing interactions and processes. Indeed, dehydration is so important in tiny biological ion channels that it is only the presence of charges, dipoles, and protein structure/dynamics that allows ions to pass by counteracting dehydration. Indeed, selectivity occurs because the pore interacts with ions in an ion-dependent manner, e.g., due to the atomic configuration in the protein's selectivity filter \cite{Doyle98-1}. Although functionalization of, and surface impurities on graphene are currently not completely known and controllable, our results suggest that graphene offers a route to characterize ion transport behavior in confined geometries without the effect of protein structure, surface charges, etc. This will likely lead to the development of experimental model systems to mimic and understand biological channels, e.g., by delineating the dehydration contribution in more complex biological systems.

Finally, one of the core challenges with the use of solid-state membranes for filtration and other applications is to engineer selectivity. Control of surface charges and site-specific chemical functionalization of, e.g., graphene, is currently not possible. Our results suggest that the geometry, which is significantly more controllable, can be designed to give selectivity. Depending on application, dehydration -- potentially with other factors -- can be exploited to control ion exclusion, selectivity, and flow rates. For instance, layering the graphene (in addition to changing the pore radius) can give measurable dehydration-only selectivities over two orders of magnitude and exclusion over many more\cite{sahu2017}. For ion separation, though, one would want to exploit dehydration or charge-based selectivity in combination with functional groups (or charge) to enhance the overall ion flow of the desired ion. Moreover, these results will shed light on the role of the electrostatic environment and functionalization introduced by the fabrication process, as well as other conditions such as pH. Together with a characterization of pore functionalization, solid-state pores will thus allow -- for the first time -- to experimentally delineate the contributions to transport in more complex biological pores, and for the optimization of porous structures for applications. \\

{\noindent \bf Methods:}
We perform all-atom molecular dynamics using NAMD2 \cite{phillips2005} and the CHARMM27 force field, where we first minimize the energy of the system and then heat it to 295 K. A 0.5 ns NPT (constant number of particles, pressure and temperature) equilibration using the Nose-Hoover Langevin piston method followed by 1.5 ns of NVT (constant number of particles, volume and temperature) equilibration  generates the initial atomic configuration. An electric bias applied perpendicular to the plane of the membrane drives the ionic current through the pore. The production run varies from 100 ns to 1.1 $\mu$s depending on convergence of the current and other properties of interest. 

We use the adaptive biasing method in the colvar module of NAMD to perform the free energy calculation. In this simulation, an ion is confined within a cylinder of height 1 nm and radius 0.2 nm centered at the origin. The biasing force helps the ion overcome potential barriers and explore the energy landscape. The total simulation consists of about 120 runs of 10 ns each to obtain the free energy difference shown in Fig. \ref{ABF}. We use the block standard error method to compute the error bars for all plots. Additional details are in the Supporting information.

{\noindent \bf Corresponding Author} \\
*E-mail: mpz@nist.gov\\

{\noindent \bf Author contributions}\\
M.Z. and M.D. proposed the project. S.S. and M.Z. designed and performed numerical calculations and developed the theory of selectivity. All authors wrote the manuscript and clarified the ideas. \\

{\noindent \bf ACKNOWLEDGMENTS}\\
S. Sahu acknowledges support under the cooperative Research Agreement between the University of Maryland and the National Institute of Standards and Technology Center for Nanoscale Science and Technology, Award 70NANB10H193, through the University of Maryland. We are grateful to S. C. O'Hern and R. Karnik for providing us the data for Ref.~\onlinecite{OHern2014}. We also thank D. Gruss, J. Elenewski,  K. Schulten, and C. Sathe for helpful comments. \\

%----------------------------------------------------------------------------------------------
%--------------------------------------------------------------------------------
%\section{REFERENCES}

%\bibliography{reference}

\end{document}

\title{Dehydration as a Universal Mechanism for Ion Selectivity in Graphene and Other Atomically Thin Pores -- Supporting Information}

\author{Subin Sahu}
\affiliation{Center for Nanoscale Science and Technology, National Institute of Standards and Technology, Gaithersburg, Maryland 20899}
\affiliation{Maryland Nanocenter, University of Maryland, College Park, Maryland 20742}
\affiliation{Department of Physics, Oregon State University, Corvallis, Oregon 97331}

\author{Massimiliano Di Ventra}
\affiliation{Department of Physics, University of California, San Diego, California 92093}

\author{Michael Zwolak}
\affiliation{Center for Nanoscale Science and Technology, National Institute of Standards and Technology, Gaithersburg, Maryland 20899}
\email{mpz@nist.gov}

\maketitle
\tableofcontents

\clearpage
\section{Methods}

\subsection{Molecular Dynamics}

Graphene has carbon atoms located at the points $\hat{r}_{nm} = m \hat{a}_1 + n\hat{r}_2$ for $m,\, n \in \mathbb{Z}$, where  $\hat{a}_1=a( 3, \sqrt{3})/2$ and $\hat{a}_2=a (3,-\sqrt{3})/2$ are the 2D lattice vectors and $a \approx 0.14$ nm is the C-C bond length. We open a pore of nominal radius $r_n$ at the center of each membrane by removing carbon atoms satisfying the condition  $x^2+y^2<r_n^2$, with $x,\, y$ the coordinates of the atom in the $z=0$ plane. We then immerse the membrane in an aqueous ionic solution, typically with 1 mol/L salt concentration, consistent with experiments. We use the CHARMM27 force field to model the atoms. The carbon atoms are type CA and water molecules are TIP3P from the  CHARMM27 force field.

We perform all-atom molecular dynamics (MD) simulations using NAMD \cite{phillips2005} with a time step of 1 fs and periodic boundary conditions in all directions. We use a cutoff of 1.2 nm for non-bonded interactions, i.e., van der Waals and electrostatics. However, we use full electrostatic calculations every 4 fs via the particle-mesh Ewald (PME) method \cite{darden1993}. We first minimize the energy of the system for 4000 steps (4 ps) and then heat it to 295 K in another 4 ps. A 0.5 ns NPT (constant number of particles, pressure and temperature) equilibration using the Nose-Hoover Langevin piston method \cite{Martyna1994} -- to raise the pressure to 101~325 Pa (i.e., 1 atm) -- followed by 1.5 ns of NVT (constant number of particles, volume and temperature) equilibration generates the initial atomic configuration. An electric field perpendicular to the plane of the membrane drives the ionic current through the pore. We set the Langevin damping rate to 0.2/ps for carbon and water (via its oxygen atoms) during these runs. Test runs show that damping the hydrogen atoms does not affect the results. Damping the ions, however, affects the current as it changes the ionic mobility. 

We fix the outer edge of the graphene membrane, but the bulk of the membrane has no confinement other than the C-C bonds of graphene. The production runs vary from 100 ns to 1.1 $\mu$s based on the convergence of the current and other properties of interest. When calculating the water density around an ion fixed at the origin (e.g., in the center of the pore when it is not fluctuating), the parameters are the same except there was no external electric field present.

We use the adaptive biasing force method (ABF) \cite{Darve2008,Henin2004} in the colvar module of NAMD to perform the free energy calculation. In this method the reaction coordinate, $\zeta$ ($z$ for the setup here), is divided into equally spaced bins and the free-energy difference along $\zeta$ is calculated by integrating the equation
\begin{equation}
 \left\langle f_\zeta \right\rangle_k = - \left\langle \frac{\partial U(X)}{\partial \zeta} - \frac{\partial ln|J|}{\partial \zeta} \right\rangle_k \equiv  - \frac{dF(k)}{d\zeta},
\end{equation}
where $\left\langle f_\zeta \right\rangle_k$ and $F(k)$ are the mean force and free energy at bin $k$, $X$ are the cartesian coordinates, and $|J|$ is the Jacobian of the transformation to cartesian coordinates. The ABF method applies an iterative biasing force, $\bar{f_n}(k) $, which is the average of all force samples after $n$ MD steps in the bin $k$.  This force enables the system to overcome free-energy barriers during an unconstrained MD run and allows for a more uniform sampling along the reaction coordinate.
In our simulation, we calculate the one-dimensional free energy profile along the $z$-axis in bins of width 0.01 nm from $z=-1.5$ nm to $z=1.5$ nm. We perform ABF calculation on three windows, ($-1.5$ nm $ \leqslant z \leqslant -0.5 $ nm), ($-0.5$ nm $ \leqslant z\leqslant 0.5$ nm), and ($0.5$ nm $ \leqslant z \leqslant 1.5 $ nm), and also symmetrize the final result about $z=0$ for better sampling. Also, we confine the ion within a cylinder of radius 0.2 nm centered at the origin, where a bounding potential with force constant $\approx 43$ eV/nm$^2$ turns on outside of the 0.2 nm cylinder.  In each bin, 800 samples of the instantaneous force are accrued prior to the full application of the ABF. This biasing force scales up linearly by a factor of 0 to 1 from 400 samples to 800 samples, and no biasing force is applied below 400 samples. 
The total simulation consists of about 120 runs of 10 ns each for each window.

We use the block standard error (BSE) method \cite{grossfield2009} to compute the error bars for all plots. The BSE is given by 
\begin{equation}
\mathrm{BSE} = \frac{s_\tau \sqrt{\tau}}{\sqrt{T}},
\end{equation}
where $T$ is the total simulation time (the time of the MD trajectory), $\tau$ is a length of time used to partition the simulation into many contiguous blocks, and $s_\tau = \sqrt{ \frac{\sum_i (\avg{I_\tau}_i -\avg{I_T})^2}{(N_b -1)}} $  is the standard deviation of the mean current, $\avg{I_\tau}$,  within each of the $N_b$ blocks. The BSE depends on $\tau$ when the latter is very small (i.e., when $\tau$ is smaller than the timescale required to get independent reads of the current) or very large ($\tau \approx T$). In the first case, the dependence is due to the fluctuations in the mean being correlated and, in the later case, the estimate of the standard deviation having too few data points. However, the BSE is fairly constant over a broad range of $\tau$ in between, which is the value we used to estimate errors. The error bars in the plots are $\pm 1$ BSE unless otherwise indicated.

\subsection{Pore radius and area}

\begin{figure}[h]
\includegraphics{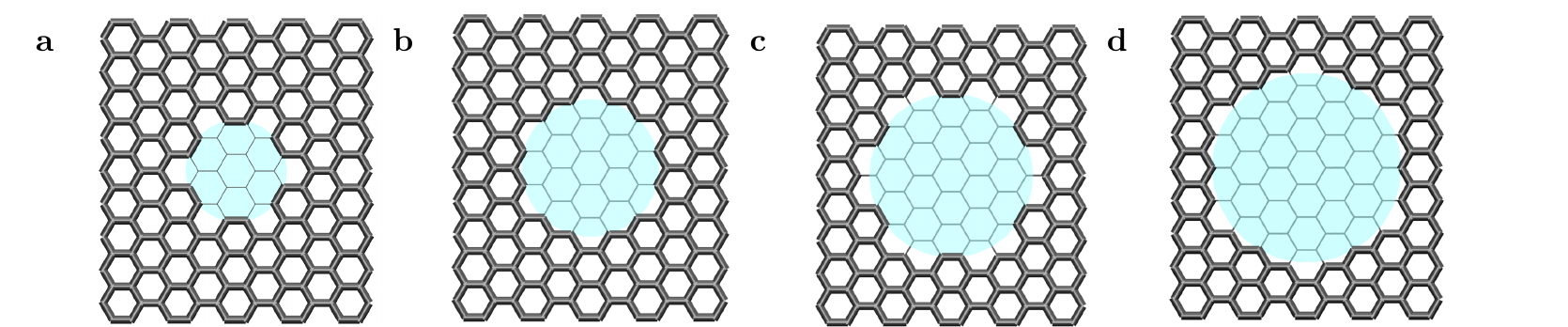}
\includegraphics{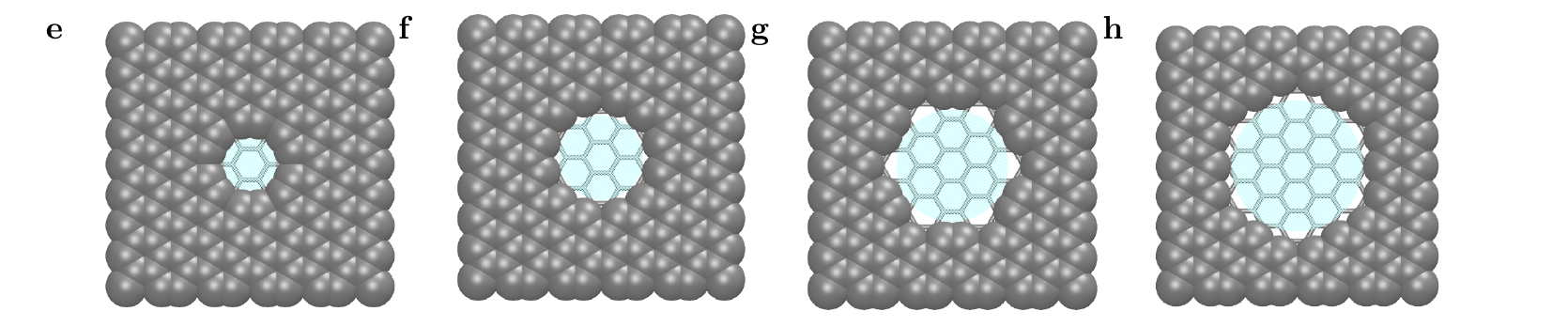}
\caption{\label{pores} Membrane and pore structure. The images show a small section of the graphene membrane (2.1 nm by 2.1 nm) containing the pore. (a-d) The top panels show the nominal radii, $r_n =$ 0.38 nm, 0.51 nm, 0.62 nm, and 0.71 nm. (e-h) The bottom panels show the corresponding effective radii, $r_p=$ 0.21 nm, 0.34 nm, 0.45 nm, and 0.54 nm.}
\end{figure}

\begin{table}[h]
\centering
\begin{tabular}{|l|P{1cm}|P{1cm}|P{1cm}|P{1cm}|P{1cm}|P{1cm}|P{1cm}|P{1cm}|P{1cm}|P{1cm}|} \hline
$r_{n}$ (nm)        & 0.38 & 0.51  & 0.62  & 0.71 & 0.86 & 0.93 & 1.11 & 1.24 & 1.36 & 1.48\\ \hline
$r_{p}$ (nm) & 0.21 & 0.34  & 0.45  & 0.54 & 0.69 & 0.76 &0.94 & 1.07 & 1.19 & 1.31\\ \hline
\end{tabular}
\caption{\label{radius} Radii for various pores. Here, $r_{n}$ is the nominal radius of the pore (it defines the construction of the pore for the simulations) and $r_{p} = r_n - \sigma_C$ is the effective radius, where $\sigma_c$ is the vdW radius of carbon. The quantity $r_{p}=r_n - \sigma_C$ also gives  the radius that would be observed in experiments due to electron density around the carbon atoms and bonds.}
\end{table} 

As noted above, we open pores by removing carbon atoms with coordinates satisfying $x^2+y^2<r_{n}^2$. There is a range of $r_n$ that give the same pore size due to the discrete nature of the membrane. We choose $r_n$ to be the maximum of this range. This $r_n$ also gives the distance of the carbon atoms at the pore edge to the center of the pore, see Fig. \ref{pores} (in other words, the radius of the largest circle that will fit into the pore). 

\begin{figure}[h]
\includegraphics{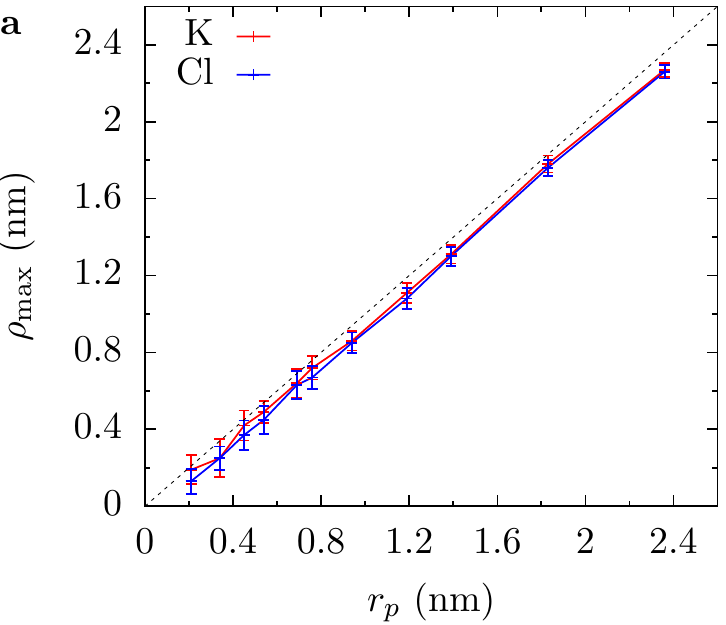} \qquad
\includegraphics{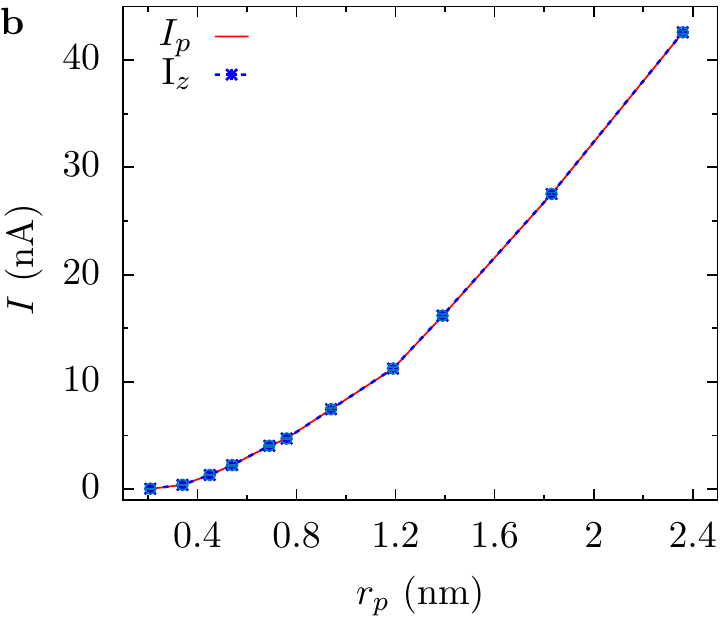} 
\caption{\label{area} Transport area and current definition. (a) The maximum (cutoff of 99 \%) radial spread ($\rho^2=x^2+y^2$) of ions inside the pore is roughly equal to $r_p$ nm. That is, when looking at the integrated density of translocation events from 0 to $\rho$, 99 \% of the events fall between 0 and $\rho_{max}$. (b) Current calculated from the two different definitions $I_p$ (solid line) and $I_z$ (dashed line). Connecting lines are shown as a guide to the eye.}
\end{figure}

When comparing to experimental results, however, these nominal radii may or may not correspond to the values reported. For instance, Ref.~\onlinecite{OHern2014} defines the pore area by where electron density is not observed in transmission electron microscopy (TEM) images. This will roughly correspond to $r_n - \sigma_C$ where $\sigma_C\approx0.17$ nm is the van der Waals (vdW) radius of carbon. This is approximately where the electron density vanishes. Moreover, the actual area available for transport is smaller than nominal area due to hydration and the finite size of atoms (note, however, that the flexibility of the pore edge will tend to slightly increase the area available). These factors have negligible effect in larger pores but are significant for subnanoscale pores. We found that in general the maximum radial spread of the ion inside the pore is given by: $\rho_{max}\approx r_n-\sigma_C$ (see Fig. \ref{area}). Thus, we define the effective pore radius as $r_p = r_n -\sigma_C$ and effective area of the pore as  $ A_p=\pi r_p^2$. This definition should roughly correspond to experimentally observed values. We note that to find the actual accessible area for transport, one should also account for the spatial dependence of the free energy. Table~\ref{radius} reports the nominal and effective radii.

\subsection{Current definition and electric field}

\begin{figure}[h!]
\includegraphics{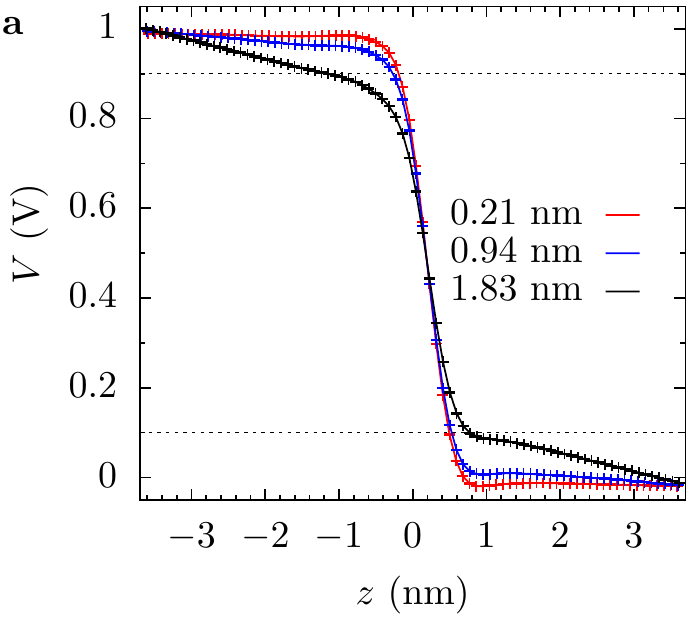}\quad
\includegraphics{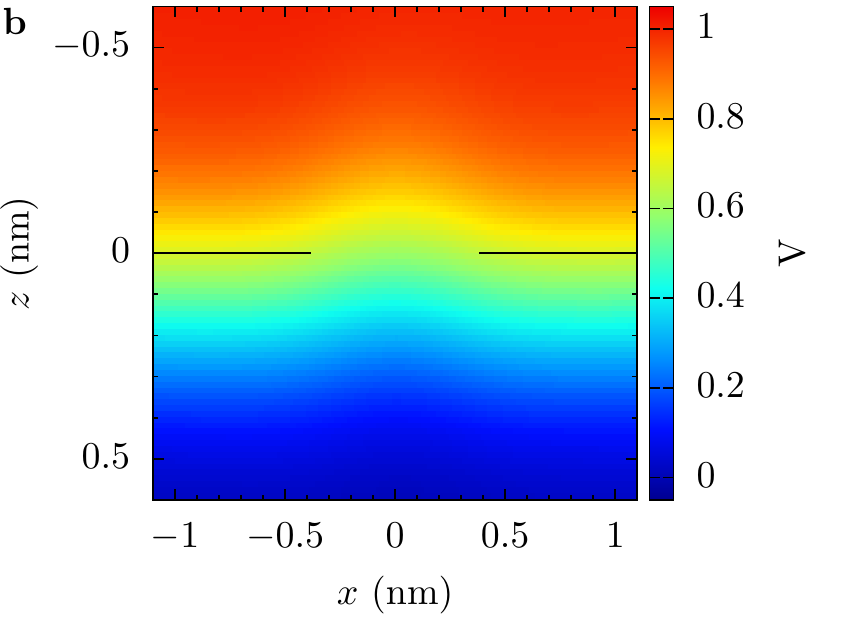}
\caption{\label{E-field} Voltage drop. (a) Potential drop along the $z$-axis for various pore sizes and (b) a map of the potential for a pore of radius 0.21 nm. For larger pores, most of the potential drop (about 80\% shown by the two dotted horizontal lines) occurs within a distance $r_p$ from the center of pore. The electric field can thus be approximated by $E \approx 0.8\, V/2r_p$. However, for small pores, the entire potential drop occurs over 1 nm due to vdW repulsion of the ions by the graphene membrane. In Fig. 2 of the main text, we use the approximate electric field for larger pores in the expression $e z_\nu n_\nu \mu_\nu A_p E$ for the whole range of values reported. This overestimates the field for smaller pores, but will approach the right value as the pore size increases. Note, as well, that $\K^+$ tends to come closer to the graphene membrane due to its smaller vdW repulsion. This results in the potential drop occurring mostly on the anion side and, when the voltage is increased enough to substantially polarize water, in the effective barrier for $\K^+$ decreasing before that for $\Cl^-$.  Connecting lines are shown as a guide to the eye. The potential maps were produced using the method described in Ref.~\citenum{aksimentiev2005}.} 
\end{figure}

The ionic current was calculated using two definitions:
\begin{equation}
I_z(t)= \frac{1}{\Delta t L_z}\sum_{i=1}^N q_i[z_i(t+\Delta t) - z_i (t)],
\end{equation}
and
\begin{equation}
I_p(t)= \frac{1}{\Delta t }\sum_{i=1}^N q_i(\Theta [Z_i(t+\Delta t)] - \Theta[Z_i (t)]),
\end{equation} 
where $\Theta$ is the Heaviside step function and $\Delta t=1$ ps is the measurement time (we record the atomic configuration every $\Delta t$ increment). The first definition takes into account the motion of all ions in the $z$-direction and the second definition counts ions crossing the pore. These definitions give the same value for the current so long as the simulation is converged with respect to the total simulation time (see Fig.~\ref{area}(b)).

We found that about 80 \% of the potential drops within a sphere of radius $r_p$ from the center of the pore for larger pores (see Fig. \ref{E-field}). Thus, the electric field inside the pore can be estimated as $E \approx 0.8\,V/2r_p$. For the smallest pore, however, the electric field is more accurately determined by $E \approx V/L$ with $L=1$ nm. This is because vdW repulsion of the ions by the carbon prevents the charge layers from getting closer than about 1 nm. 

We tested the effect of box size on the current by comparing different box sizes, a large box (with fixed cross-sectional area 7.4 nm $\times$ 7.4 nm and relaxed height 6.9 nm), a small box (with fixed cross-sectional area 3.7 nm $\times$ 3.7 nm and relaxed height 3.4 nm), and a extended small box (with fixed cross-sectional area 3.7 nm $\times$ 3.7 nm and relaxed height 6.9 nm). The current and selectivity are in agreement (to within the errors reported) for all box sizes when the pore is small and at low voltages, as the current is dominated by the high pore resistance. The latter allows well-defined charge layers to develop and persist. However, for larger pores, the current was smaller by about 20 \% for the larger box size. Since both box sizes have the same voltage and same concentration of ions, the difference in current is likely due to a smaller access resistance and lack of well-defined charge layers in the boxes with smaller cross-sectional area, as the pore diameter approaches the edge of the box. For the smallest pore size calculations, we use the extended small box, as it considerably reduces errors due to convergence in time, i.e., we can run microsecond long simulations, which are necessary when the currents are so small. 

\begin{figure}[h]
\includegraphics{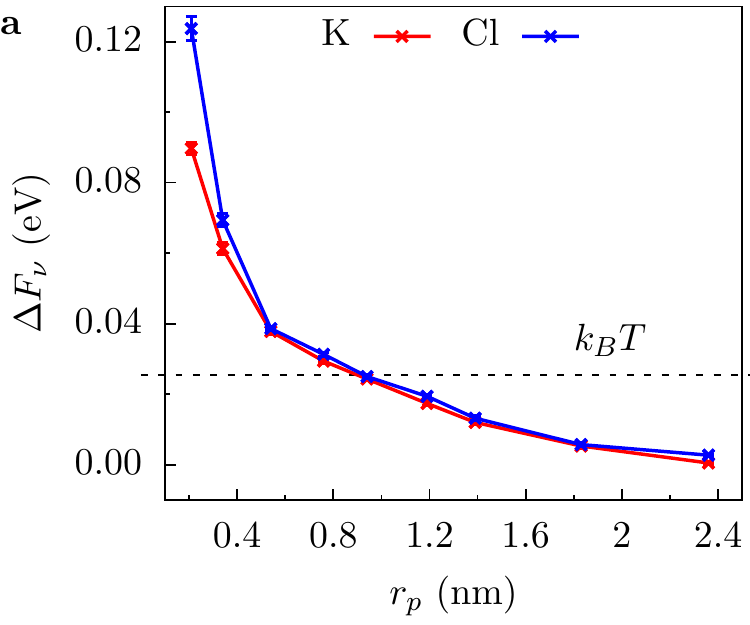} \quad
\includegraphics{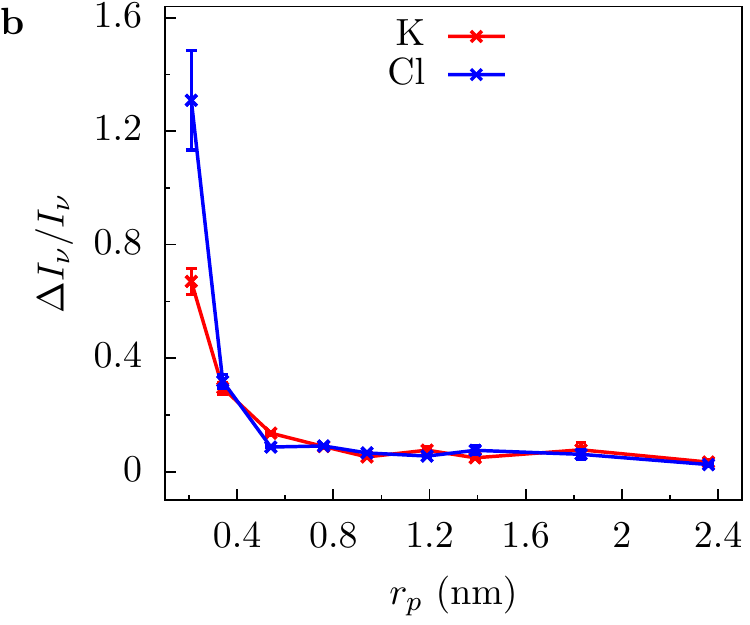}
\caption{\label{noise} Estimates of the free energy barrier and noise for various pores. Sharp rise in (a) the free energy barrier and (b) the noise in the current at the subnanoscale. An ``average'' free energy barrier is estimated as $\Delta F_\nu = k_{B} T\, \log\left(\frac{q_\nu n A_p \mu_\nu E}{I_\nu}\right)$, where $q_\nu$, $\mu_\nu$, $n$ are the charge, average mobility (of $\K^+$ and $\Cl^-$), and particle concentration. The error in $\Delta F_\nu$ is thus $k_{B} T \delta I_\nu/I_\nu$, where $\delta I_\nu$ is BSE in $I_\nu$. The noise in current, $\Delta I_\nu(\tau)$ is measured by standard deviation of the mean current within blocks of length $\tau=10$ ns. The relative noise in the current, $\Delta I_\nu(\tau) / I_\nu$  increases sharply due to the fluctuations in hydration layer configurations that  allow or prohibit passage of ions through the pore \cite{zwolak10-1}. The errors in $\Delta I_\nu$ is estimated as $\frac{\Delta I_\nu}{\sqrt{2(N_b-1)}}$  \cite{ahn2003}. Connecting lines are shown as a guide to the eye.}
\end{figure}

\section{Current behavior at the subnanoscale}
The ionic current begins to show abnormal behavior as the radius of the pore decreases to the sub-nanometer scale. This behavior is seen in the sharp rise in the pore resistance and noise in the current, Fig. \ref{noise}. At these length scales, the pore edge begins to deform the hydration layers around the ions, which increases the energy barrier for ions to cross the pore. \\

\section{Selectivity}

\subsection{Experimental observation of selectivity}

\begin{table}[]
\centering
\begin{tabular}{|l|P{2.25cm}|P{2.25cm}|P{2.25cm}|P{2.25cm}|P{2.25cm}|} \hline
$A_p$ (nm$^2$)        & 0.2 & 0.123  & 0.108  & 0.321 & 0.368\\ \hline
$r_{p}$ (nm) & 0.25 & 0.20  & 0.19  & 0.32 & 0.34\\ \hline
$S_{p}$ & S & S  & S  & 1 & 1\\ \hline
\end{tabular}
\caption{\label{selectivity-table}Current-carrying pores from the selectivity measurement of Ref.~\citenum{OHern2014}. We assign a selectivity $S$ for all pores with $r_p\approx 0.2$ nm and 1 for larger pores. There are also many pores with smaller radii, but we expect these pores to carry negligible current (and thus, even though they will be selective, they do not contribute to the observed selectivity). We do not include them in this table.}
\end{table}

\begin{table}[h!]
\centering
\begin{tabular}{|l|P{2cm}|P{2cm}|P{2cm}|P{2cm}|P{2cm}|} \hline
Salt (XCl) &    CsCl & RbCl& KCl& NaCl & LiCl \\ \hline
$\sigma$ (pS) & 67 & 70 & 64& 42& 27\\ \hline

$\kappa$ (10$^{-3}$ S m$^{-1}$) &1.42 & 1.42 &1.3 & 1.19 &0.95 \\ \hline
$E_\mathrm{X}$ (eV) &3.1 & 3.4 &3.7 & 4.6 &5.7 \\ \hline
$\mu_\mathrm{X}$ (10$^{-8}$ m$^2$V$^{-1}$s$^{-1}$) & 8.01 &8.06  &7.62 & 5.19 &4.01 \\ \hline
$\mu_\mathrm{eff}$ (10$^{-8}$ m$^2$V$^{-1}$s$^{-1}$) & 4.25 &3.59  &4.29 & 1.63 &1.03 \\ \hline
$\sigma_\mathrm{XCl}$ (pS) &67 & \tr{61}-\tb{67} &\tb{65}-\tr{67}& \tr{44}-\tb{55} & \tr{39}-\tb{50} \\ \hline
$\sigma_\mathrm{X}$ (pS) &67 &\tr{57}-\tb{66} & \tb{64}-\tr{68} & \tr{26}-\tb{43} &\tr{16}-\tb{34} \\ \hline
\end{tabular}
\caption{\label{garaj} Leakage conductance ($\sigma$), bulk conductivity of cations and anions together ($\kappa$), and hydration energy ($E_\mathrm{X}$) of the cation $\mathrm{X}$, as reported by Ref.~\citenum{garaj2010}. Note that the reported bulk conductivity is four orders of magnitude smaller than that in their nanopore current measurements. We take the cation mobilities ($\mu_\mathrm{X}$) from bulk~\cite{hille2001} and effective cation mobilities ($\mu_\mathrm{eff}$) in a biological pore from Ref.~\citenum{bhattacharya2011}. The bulk mobility of $\Cl^-$ ion is $7.92 \times 10^{-8}$ m$^{-2}$ V$^{-1}$ s$^{-1}$ and its effective pore mobility (from Ref.~\citenum{bhattacharya2011}) is $3.44 \times 10^{-8}$ m$^{-2}$ V$^{-1}$ s$^{-1}$. We make two estimates of the range of the conductance: One for $\sigma_\mathrm{XCl}$ (assuming both the cation and anion contribute) and the other for $\sigma_\mathrm{X}$ (assuming only the cation contributes). The range in each case is set by the bulk mobility and the effective pore mobility. We note that even though graphene is atomically thick (and effective mobilities are not well defined), the defect channel structure is not known -- those channels could be long channels through gaps in the device. We use the effective mobilities of Ref.~\citenum{bhattacharya2011} for a biological pore only as a very rough estimate. These ranges show that the deviation of the relative conductance from that predicted by bulk mobilities can easily be due to cation-only conductance (e.g., due to local charge-based selectivity) and/or effective mobilities through the defect channels responsible for the leakage conductance. Note that the conductance of CsCl and Cs only in the last two lines are the same. This is because they are assuming two different hypotheses about the origin of the conductance and we take them both to be the experimentally determined conductance. Also, for clarity, the estimated range is color coded with blue indicating the estimate using the bulk mobility and red the estimate from the effective mobility.}
\end{table}

\begin{table}[h!]
\centering
\begin{tabular}{|l|P{2cm}|P{2cm}|P{2cm}|P{2cm}|P{2cm}|} \hline
cation(X) &    K$^+$ & Li$^+$& Ba$^{2+}$& Ca$^{2+}$ & Mg$^{2+}$ \\ \hline
$\mu_\mathrm{X}$ (10$^{-8}$ m$^2$V$^{-1}$s$^{-1}$)  & 7.62 & 4.01 & 6.60& 6.17& 5.50\\ \hline
&\multicolumn{5}{c|}{device 3, $r_p=(0.36\pm 0.10)$ nm} \\ \hline
$\sigma$ (nS) & 2.3 $\pm$ 1.2& 1.1$\pm$ 1.1& 1.0$\pm$0.8& 0.7$\pm$ 0.5&0.4$\pm$ 0.03\\ \hline
$S_\mathrm{XCl/KCl}$& 1    & 0.00 - 2.61    & 0.06 - 1.75    & 0.06 - 1.25    & 0.11 - 0.41    
  \\ \hline
$S_\mathrm{X/K}$ & 1    & 0.00 - 3.80    & 0.07 - 1.89    & 0.07 - 1.40    & 0.13 - 0.49    
 \\ \hline
&\multicolumn{5}{c|}{device 4, $r_p=(0.50\pm 0.10)$ nm}  \\ \hline
$\sigma$ (nS)& 4.2 $\pm$ 0.3&2.2$\pm$ 1.3& 1.4 $\pm$0&1.5$\pm$0.1& 1.3$\pm$ 0.1\\ \hline
$S_\mathrm{XCl/KCl}$ & 1 & 0.25 - 1.17    & 0.34 - 0.40    & 0.34 - 0.45    & 0.31 - 0.39    
 \\ \hline
$S_\mathrm{X/K}$ & 1    & 0.37 - 1.71    & 0.37 - 0.43    & 0.38 - 0.51    & 0.37 - 0.46    
 \\ \hline
&\multicolumn{5}{c|}{device 8, $r_p=(0.39\pm 0.06)$ nm} \\ \hline
$\sigma$ (nS)& 2.6$\pm$1.5 &1.3$\pm$0.1 & 2.3$\pm$1 &2$\pm$1 &1$\pm$1 \\ \hline
$S_\mathrm{XCl/KCl}$   & 1& 0.39 - 1.65    & 0.32 - 3.21    & 0.28 - 3.01    & 0.00 - 2.00
 \\ \hline
$S_\mathrm{X/K}$& 1    & 0.57 - 2.42    & 0.35 - 3.46    & 0.31 - 3.37    & 0.00 - 2.39    
 \\ \hline
\end{tabular}
\caption{\label{Jain} Chloride salt conductance ($\sigma$) of three different devices holding the chloride concentration constant at 100 mM~\cite{Jain2015}. The quantity $S_\mathrm{XCl/KCl}$ is the selectivity quantified by assuming both cation and anion contribute, Eq. \ref{SXCl}, and $S_\mathrm{X/K}$ by assuming only cations contribute, Eq. \ref{SX}. The selectivity is shown as a range based on the error in $\sigma$ (the actual range -- the range of the data measured in Ref.~\citenum{Jain2015} -- is larger than shown here). Only bivalent ions in device 4 and Mg$^{2+}$ in device 3 potentially show selective behavior. However, selectivity of this magnitude was observed in Ref.~\citenum{rollings2016} for large pores, where dehydration can not be playing a role.}
\end{table}

Ref.~\onlinecite{OHern2014} found that graphene membranes with a distribution of pore sizes (at the subnanoscale) display selective behavior for K$^+$ over Cl$^-$. In that work, this is indicated by a nonzero membrane potential $E_m$. We use the relation 
\begin{equation}
E_m=\frac{k_B T}{e} \ln \left(\frac{P_{\K}[\K]_o+P_{\Cl}[\Cl]_i}{P_{\K}[\K]_i+P_{\Cl}[\Cl]_o}\right)
\end{equation}
or, rearranging, 
\begin{equation}
\quad e^{e E_m/k_B T} =\frac{\frac{P_{\K}}{P_{\Cl}}[\K]_o+[\Cl]_i}{\frac{P_{\K}}{P_{\Cl}}[\K]_i+[\Cl]_o}
\end{equation}
to estimate the selectivity from the reported membrane potential ($E_m = 3.3$ mV $\pm 1 $ mV, where we keep the second digit to not introduce rounding error,   $[\K]_i = [\Cl]_i =  0.17$ mol/L,   $[\K]_o = [\Cl]_o = 0.5$ mol/L , $T = 297$ K, and $e$ is the magnitude of the electron charge). We obtain $\tilde{S}=P_{\K}/P_{\Cl}=1.3  \pm 0.1$  as an ``average'' selectivity for their distribution of pore sizes. The ratio $P_{\K}/P_{\Cl}$ is the concentration-imbalance equivalent of $I_{\K}/I_{\Cl}$. 

To compare with our numbers, we need to extract the selectivity for particular pore sizes or to use the MD results to compute the average for the distribution of pore sizes in experiment. We will do both. We first note that the average selectivity from experiment is the same as the selectivity we find for $r_p=0.34$ nm (see Fig. 2 in the main text and Table~\ref{current-ratio}). Although the experimental selectivity is for the distribution of pore radii from $r_p\approx 0.1$ nm to $r_p\approx 0.3$ nm, it is likely that most of the current and hence the average selectivity is dominated by pores with $r_p \approx 0.3$ nm. In fact, we find that current for $r_p=0.34$ nm is an order of magnitude larger than that for $r_p=0.2$ nm, supporting that the slightly larger pore may be dominating the average.

To go further, though, we will first use a rough estimate to extract the selectivity for the $r_p \approx 0.2$ nm from experiment and then separately show that, when using the areas and free energies from our MD calculations, we get an average selectivity similar to experiment. Both of these calculations confirm that dehydration-only selectivity yields results in agreement with experiment. 

For a membrane with a distribution of pore sizes, the observed selectivity $\tilde{S}$ can be roughly estimated as
\begin{equation}\label{first}
\tilde{S} =\frac{\sum_{p} I_{pK}} {\sum_{p} I_{pCl}}=\frac{\sum_{p} S_{p} I_{pCl}} {\sum_{p} I_{pCl}}\approx\frac{\sum_{p} S_{p} A_p }{\sum_{p}A_p}
\end{equation}
or
\begin{equation}\label{second}
\tilde{S} =\frac{\sum_{p} I_{pK}} {\sum_{p} I_{pCl}}=\frac{\sum_{p}  I_{pK}} {\sum_{p} (1/S_{p})I_{pK}}\approx\frac{ \sum_{p} A_{p}} {\sum_{p} (1/S_{p}) A_p},
\end{equation}
where $S_{p}$ ($A_{p}$) is the selectivity (area) for a pore of radius $r_p$ and the sum over $p$ goes over individual pores. The approximate expressions in each equation assume that the current is proportional to area, but without a free energy barrier (see below for the average computed with free energy barriers). This is a strong approximation. It requires, at the least, that the smallest pore sizes (i.e., with radii of about 0.15 nm and below) to be dropped from the sum (as their current contribution is negligible and not necessarily proportional to area). The sum is thus over the data in Table \ref{selectivity-table}, which is from the selective membrane (5 min etch time) of Ref.~\onlinecite{OHern2014}. This gives $S=1.8\pm 0.3$ from Eq.~\eqref{first} and $S=2.5\pm 1$ from Eq.~\eqref{second}, where we again keep the second digit to not introduce rounding error. The difference in the selectivity estimated from the two equations is due to the fact that they did not take into account the free energy barrier at smaller radii. The error is calculated based on the error in $\tilde{S}$ only, as the error in area is not reported in Ref.~\onlinecite{OHern2014}. However, the error due to area is expected to be much smaller than the approximations in the equations. In the main text, we report the range $1.8$ to $2.5$ for the selectivity of the $r_p \approx 0.2$ nm pores.

A more accurate calculation requires the value of free energy barrier and selectivity for each of the pores in the distribution. We can, however, estimate the average selectivity for their pore size distribution (Table~\ref{selectivity-table}) using the free-energy barriers in Fig~\ref{noise}, 
\begin{equation}\label{averageSelectivity}
\tilde{S}_\mathrm{estimate} =\frac{\sum_{p} I_{pK}} {\sum_{p} I_{pCl}}=\frac{\sum_{p} \mu_\K\, A_p  e^{-\Delta F_\mathrm{K}/K_BT}_{p} }{\sum_{p}\mu_\Cl\, A_p e^{-\Delta F_\mathrm{Cl}/K_BT}_{p} }\approx 1.5,
\end{equation}
which is very close to the $\approx 1.3$ result from experiment. Here, we again dropped the smallest pores (ones smaller than $r_p \approx 0.2$ nm) from the average, as the free energy barrier will be more substantial and suppress their contribution to the selectivity.

We note a few important limitations of this comparison between the experiments and our calculations. The lowest voltage in our simulations is still order of magnitude larger than the equivalent chemical potential difference in the experiments. The regime below 0.25 V is very difficult to reach using all-atom MD simulation for these pore sizes. Moreover, current experimental techniques cannot control the functionalization of graphene nanopores (and functionalization/surface species of the graphene membrane) which depends on the fabrication method and other factors. Since it is not clear what groups will be present and where, we choose the simplest pore -- the one with no functionalization. As well, some functionalization will introduce charges/dipoles to the pore region. However, if these are close (either on the pore rim or nearby), they will not give rise to the weak selectivity observed, but rather a strong selectivity, unless the charge is very small in magnitude.

Garaj et al.~\cite{garaj2010} report that the leakage current (the current through a graphene membrane when the pores have yet to be constructed) deviates from what the bulk conductivity would predict, which they conjecture may be due to dehydration. Taking the experimental conductance of CsCl to be the reference value, we can estimate the conductance of other salts XCl as
\begin{equation}\label{sigmaXCl}
\sigma_\mathrm{XCl}=\frac{\sigma_\mathrm{CsCl}}{(\mu_\mathrm{Cs}+\mu_\mathrm{Cl})}(\mu_\mathrm{X}+\mu_\mathrm{Cl}). 
\end{equation} 
As shown in Table \ref{garaj}, the difference in {\em effective} mobilities of ions inside of pores explains some of the deviation. In fact, since the leakage conductance varies widely from membrane to membrane (by a factor of two \cite{garaj2010}), different effective mobilities alone may explain the deviation to within experimental uncertainties. 

An alternative (or complementary) explanation is that the defect channels -- the structure of which is unknown -- are cation selective due to the presence of negative (partial) charges. In this case, the estimated conductance is
\begin{equation}\label{sigmaX}
\sigma_\mathrm{X}=\frac{\sigma_\mathrm{Cs}}{\mu_\mathrm{Cs}}\mu_\mathrm{X}. 
\end{equation} 
This estimate accurately explains the observations by itself. Relying on dehydration requires a convoluted explanation -- or minuscule dehydration, e.g., a 1/100$^\mathrm{th}$ fractional removal of water -- to account for the differences between K$^+$, Na$^+$, and Li$^+$ when comparing to Cs$^+$. For instance, Na$^+$ and Li$^+$ have a much larger hydration energies than Cs$^+$, Rb$^+$, or K$^+$, and thus one expects that, if dehydration is a factor, the membrane conductance would go from cation and anion both contributing (for Cs$^+$, Rb$^+$, and K$^+$) to just anion contributing (for Na$^+$ and Li$^+$). In other words, NaCl and LiCl would have the same conductance, but they differ by 50 \%.

We also note that ion transport for different cations was measured in Ref.~\onlinecite{Jain2015}, where the authors claim to have observed hydration-based selectivity. However, their selectivity can be explained based on difference in cation mobilities. Selectivity with respect to KCl can be captured via the normalized conductance  
\begin{equation}\label{SXCl}
S_\mathrm{XCl/KCl}=\frac{\sigma_\mathrm{XCl}/(\mu_\mathrm{X}+\mu_\mathrm{Cl})}{\sigma_\mathrm{KCl}/(\mu_\mathrm{K}+\mu_\mathrm{Cl})}.
\end{equation} 
As shown in Table~\ref{Jain}, the selectivity has a large range due to a large variation of the conductance. This makes it difficult to determine if there is any selectivity at all. Only the bivalent ions in device 4  and Mg$^{2+}$ in device 3 have the entire range of $S_\mathrm{XCl/KCl}$ less than 1.  Moreover, the experimental results seem to indicate the presence of negative charges near the pore (see Fig. 3d in Ref.~\onlinecite{Jain2015}), which would mean that most of the current is carried by the cations. Thus, selectivity should be quantified as
\begin{equation}\label{SX}
S_\mathrm{X/K}=\frac{\sigma_\mathrm{XCl}/\mu_\mathrm{X}}{\sigma_\mathrm{KCl}/\mu_\mathrm{K}}.
\end{equation} 
This shows that~\eqref{SXCl} overestimates the deviation from non-selective behavior. As we mention in the main text, even if one ignores the large variation in the measured conductance in individual devices (and from device to device), the selectivity reported is consistent with the charge-based selectivity observed in Ref.~\onlinecite{rollings2016}. Those latter results were in pores much bigger than hydrated ions, indicating that the selective behavior is likely due to differences in how bivalent versus monovalent ions screen the charged pore/membrane. We finally note that the difference in hydration energy of monovalent and divalent ions is around 10 eV. It seems unlikely that this would give selectivity on the order of a factor of 2.

\begin{figure}[h!]
\includegraphics{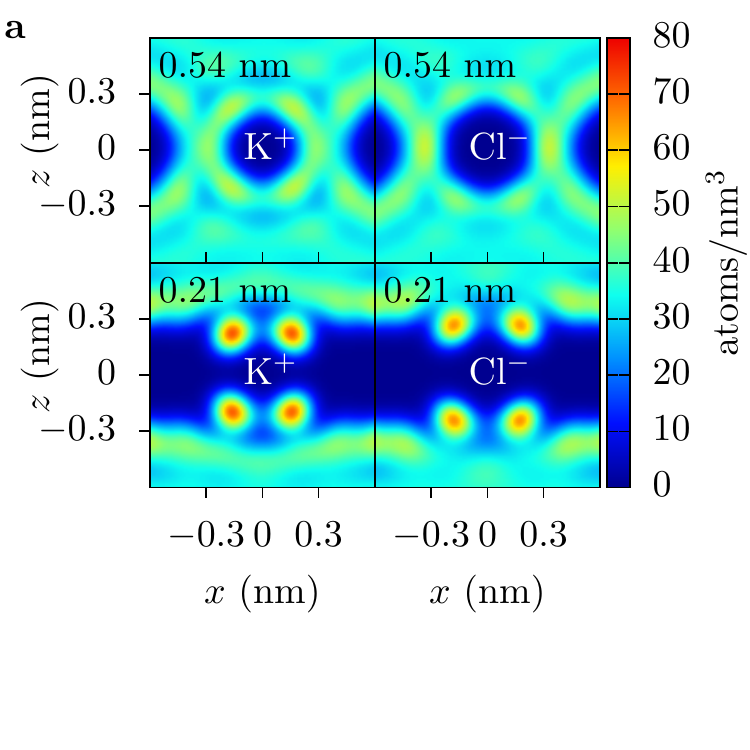} \quad
\includegraphics{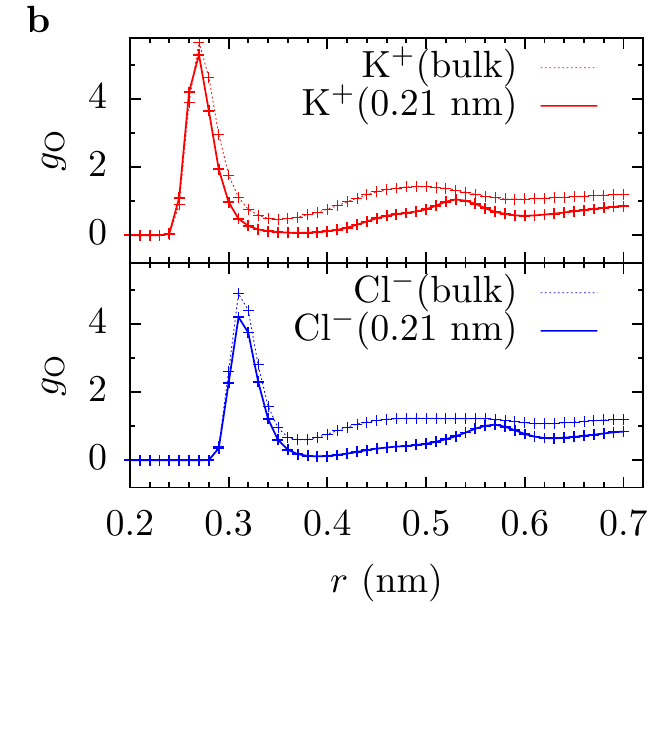}
\vspace*{-1cm}
\caption{\label{hydration} Water density quantified by the oxygen density. (a) The contour plots show either a potassium or a chloride ion fixed at the center of the pore. The pore radius is shown in the upper left corner of each panel. The hydration layers are visible as the high-density region around the ion. As the pore size gets smaller, the hydration layers around the ion are distorted by the pore edge. The ions remain fairly well hydrated. (b) The radial distribution function of oxygen atoms around $\K^+$ and $\Cl^-$ in bulk and at the center of a $r_p=0.21$ nm pore (with a counter ion fixed at the edge of the simulation cell) for zero bias showing partial dehydration. The number of water molecules in first hydration layer of $\K^+$ and $\Cl^-$ are  6.8 and 7.4 in bulk and 5.2 and 5.8 at the center of pore, respectively, i.e., a loss of about 1/4 of the water molecules from first hydration layer.  Connecting lines are shown as a guide to the eye.}
\end{figure}

\subsection{Dehydration and selectivity}

Figure \ref{hydration} shows the water density around the ion in the pore and the radial distribution function of oxygen, $g_O$.  In our simulation, the number of oxygen atoms in first hydration layer around $\K^+$ and $\Cl^-$ in bulk water are 7.4 and 6.8, respectively, and, when placed in the center of a $r_p=0.21$ nm pore, 5.8 and 5.2. This is loss of about 1/4 of the water molecules from the innermost layer. Even though the fractional dehydration for $\K^+$ and $\Cl^-$ are nearly equal, the energy penalty is higher for Cl$^-$ as its inner hydration layer is more strongly bound compare to K$^+$. As we increase the applied voltage, both ions were able to remain more hydrated while crossing the pore due to polarization-induced chaperoning of the ions. This results in a lower free energy barrier for transport, as seen from Fig. \ref{IV-model}.

\section{Model for ion transport}

\begin{figure}[h!]
\includegraphics{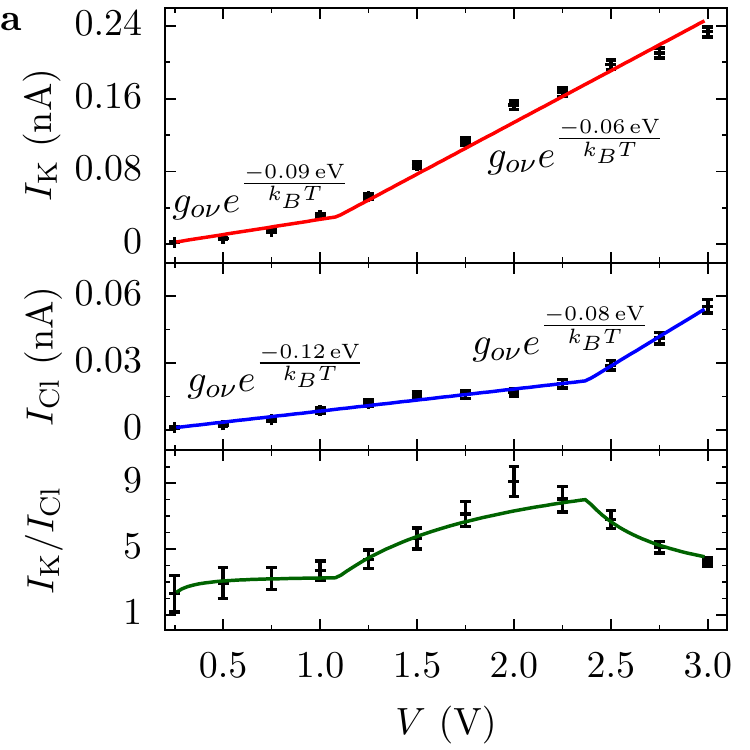}
\includegraphics{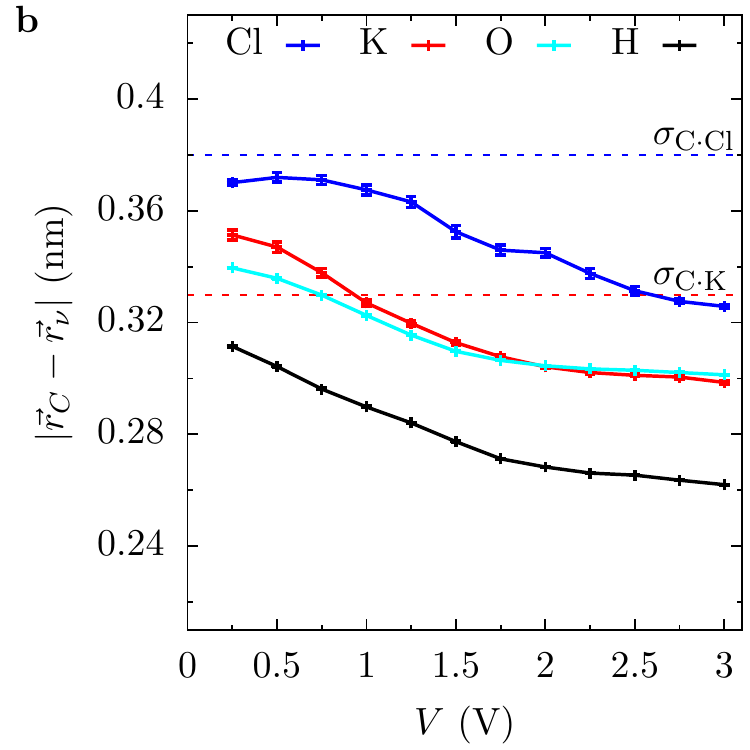}
\caption{\label{IV-model} Model for ion transport. (a) Current-voltage characteristics for a pore of radius 0.21 nm and 1 mol/L KCl versus voltage (top panels). The relative selectivity of $\K^+$ over $\Cl^-$ in the same pore (bottom panel).  Data points give the MD results and the solid line gives the piece-wise linear model we fitted taking each region to be linearly related to the differential conductance $g_\nu=e z_\nu n_\nu e^{(-\Delta F_\nu/k_B T)} \mu_\nu A_p/L = g_{0\nu} e^{(-\Delta F_\nu/k_B T)}$. Around $(1.1\pm0.1)$ V the energy barrier for potassium drops from $(0.09\pm 0.004)$ eV to $(0.06\pm0.002)$ eV, whereas that of chloride  drops from $(0.12\pm 0.001)$ eV to $(0.08\pm0.008)$ eV around $(2.38\pm0.04)$ V. This change in conductance with voltage results in a rise and fall of selectivity. (b) The distance ($|\vec{r}_C-\vec{r}_\nu|$) between an ion $\nu$ (or an oxygen/hydrogen from water) crossing the pore and the nearest carbon atom versus voltage. At higher voltage ions are able to apparently enter the repulsive zone ($r<\sigma$) of vdW potential of carbon atom, likely due to larger forces that can take advantage of the flexibility of the pore rim. This effectively increases the area of transport. This is partly responsible for the increased current at higher voltages. However, the main cause of the latter is the chaperoning of ions across the pore by polarized water, which happens at lower voltage for K$^+$ compare to Cl$^-$ due to the charge layer of the former being closer to the graphene membrane, see Fig.~\ref{E-field}(a).} 
\end{figure}

\begin{figure}[h!]
\includegraphics{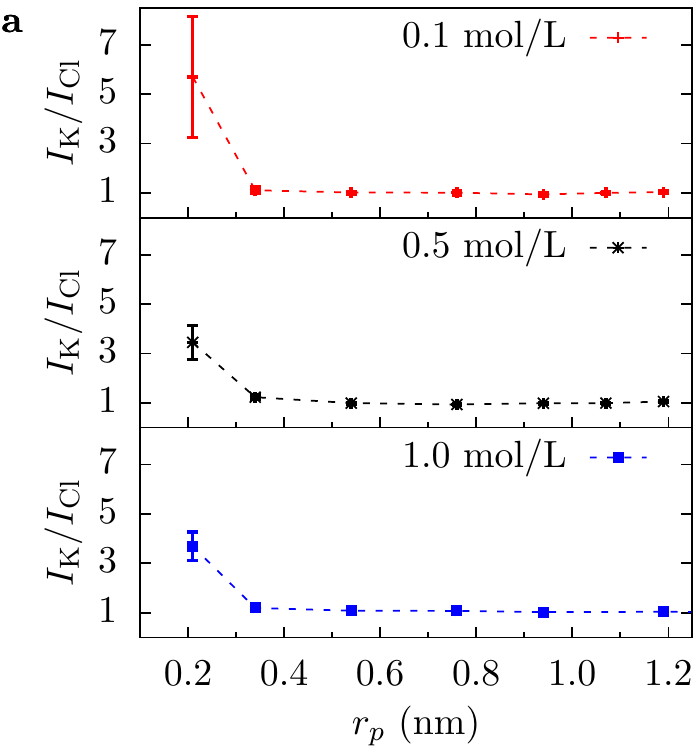}\quad
\includegraphics{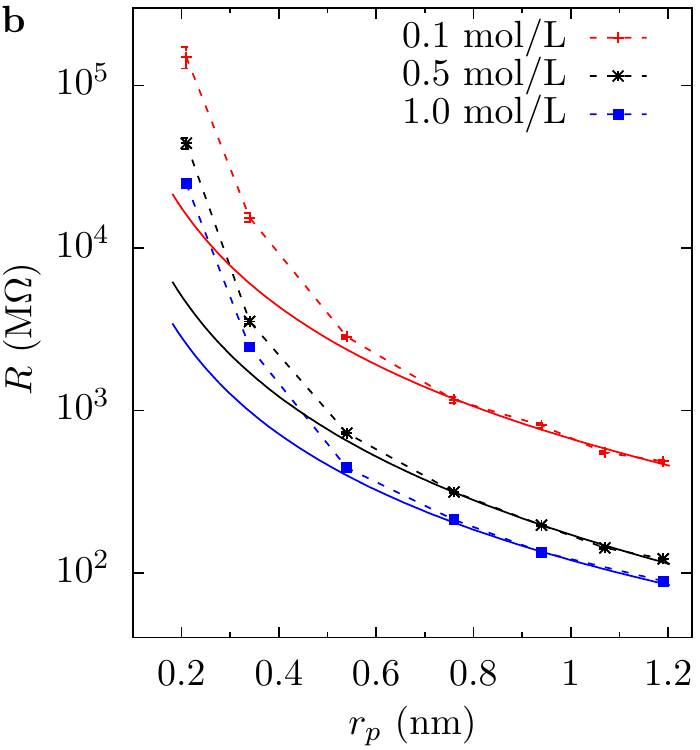}
\caption{\label{current-ratio} Selectivity and pore resistance for three different concentrations. (a) Selectivity and (b) pore resistance for 0.1 mol/L, 0.5 mol/L, and 1.0 mol/L KCl.  The solid line is the fit of resistance of the form, $R=\frac{a}{r_p}+\frac{b}{r_p^2} $ \cite{Hall1975,Schneider2010}. Here, the 1/$r_p$ term is due to access resistance and the 1/$r_p^2$ term is due to the pore resistance. Note that this deviation from normal behavior is quite large as the axis is on a logarithmic scale.  The persistence of selectivity and the sharp rise in pore resistance at lower concentrations confirms that the behavior is not due to ion-ion interactions or some other many-body effect.  Connecting lines are shown as a guide to the eye.} 
\end{figure}

The model for ion transport is discussed in detail in the main text. We use the method of least squares, with the first data voltage point constrained to reflect behavior at lower voltage, to fit the data. We only present data up to 1.5 V in the main text as water starts to dissociate at high fields. However, all-atom MD simulations allow us to apply much higher voltages without dissociating water. We thus looked at the IV characteristics up to 3 V to check consistency with our model, which fits well up to that voltage when accounting for a change in free energy barrier for $\Cl^-$, as seen in Fig. \ref{IV-model}.

%\clearpage
\section{The effect of the ion concentration}
In order to confirm that the anomalous behavior of current and weak selectivity is not due to a many-body effect but rather to single ion behavior, we repeat the selectivity calculations for lower concentrations of KCl. Both the anomalous behavior of current and weak selectivity are still present in concentrations as low as 0.1 mol/L, at which point there were only a few ions of each type in the smallest simulation box size (hence, box size errors start to become significant).

\clearpage
\section{Tables}

\begin{table}[h]
\centering
\begin{tabular}{| l|P{1.1cm}|P{1.1cm}|P{1.1cm}|P{1.1cm}|P{1.1cm}|P{1.1cm}|P{1.1cm}|P{1.1cm}|P{1.1cm}|} \hline
$r_p$ (nm)  &  0.21 &  0.34  &  0.54  &  0.76 &  0.94 &  1.19 &  1.39 &  1.83 &  2.36\\  \hline 
$l$ (nm)  &  3.7 &    3.7 &  3.7 &  3.7 &  3.7 &  7.4 &  7.4 &  7.4 &  7.4\\  \hline 
$I_\K$ (nA)& 0.032         & 0.23         & 1.13         & 2.40         & 3.57         & 5.7
        & 8.1         & 13.6        & 21.8 \\ \hline
$I_\Cl$ (nA) & 0.009         & 0.17         & 1.13         & 2.32         & 3.61         & 5.5
        & 8.0         & 13.9        & 20.8 \\ \hline
$I_\K /I_\Cl$ & 3.7         & 1.3         & 1.0         & 1.0         & 1.0         & 1.0
        & 1.0         & 1.0         & 1.0\\ \hline
\end{tabular}
\caption{\label{radius} Current for various radii. The table shows the current for 1.0 mol/L KCl solution for the box of height $h=6.9$ nm for various radii. The edge of the box is $l=b=7.4$ nm for larger pores ($r_p>1$ nm) and $l=b=3.7$ nm for smaller pores. The latter allows for much longer simulations, which are needed to achieve convergence. The error in the current is shown in Fig. 2 in the main text. The block standard error determines the number of significant digits in this and the following tables.}
\end{table}

\begin{table}[h!]
\centering
\begin{tabular}{| l|P{.75cm}|P{.75cm}|P{.75cm}|P{.75cm}|P{.75cm}|P{.75cm}|P{.75cm}|P{.75cm}|P{.75cm}|P{.75cm}|P{.75cm}|P{.75cm}|} \hline
$V$ (V)        &0.25 & 0.5 & 0.75   & 1.0 & 1.25 & 1.5 &1.75 & 2.0 & 2.25 &2.5 &2.75 &3.0\\  \hline 
$I_\K$ (pA) &2.3 &    7.0 &    15 &    32 &    52 &    87 &    113 &    153 &    167 &    197 &    210 &    234  \\ \hline
$I_\Cl$ (pA) & 1.0      & 2.2      & 4.5      & 9      & 12      & 15      & 16      & 17      & 21      & 29      & 41      & 55       \\ \hline
$I_\K /I_\Cl$ & 2.3      & 3.0      & 3.2      & 3.7      & 4.4      & 5.6      & 7.1      & 9.1      & 8.0      & 6.8      & 5.1      & 4.2 \\ \hline
\end{tabular}
\caption{\label{voltage} Current for various voltages. The table shows the current for 1.0 mol/L KCl solution through the smallest pore $r_p=0.21$ nm for various voltages. For these calculations, we use the box height $h=6.9$ nm and the smaller cross section, $l=b=3.7$ nm. The error in the current is shown in Fig. \ref{IV-model}}
\end{table}

\begin{table}[h]
\centering
\begin{tabular}{|P{1.75cm}|P{1.75cm}|P{1.75cm}|P{1.75cm}|P{1.75cm}|}
 \hline
 \multirow{2}{*}{}  & \multicolumn{2}{c|}{pore} & \multicolumn{2}{c|}{bulk} \\ \cline{2-5}
   & $\avg{p_r}$ (D)    & $\avg{n}$                & $\avg{p_r}$ (D)                 & $\avg{n}$    \\ \hline
K$^+$  & 2.02 & 5.2               & 1.60               & 6.8  \\ \hline
Cl$^-$ & 1.81 & 5.8               & 1.54               & 7.4  \\ \hline
\end{tabular}
\caption{\label{Dipole} Dipole orientation. The average radial component of individual water dipole $\avg{p_r}$ and average number of water dipoles $\avg{n}$ in the first hydration layer of K$^+$ an Cl$^-$ ions in the smallest pore ($r_p=0.21$ nm) and in the bulk. The dipole moment of water in our model is 2.35 in units of Debye (0.021 e nm). The dipoles are strongly oriented in the pore compare to bulk but have fewer dipoles and hence there is a decrease in total dipole moment.}
\end{table}

\begin{table}[h!]
\centering
\begin{tabular}{| l|P{1.2cm}|P{1.2cm}|P{1.2cm}|P{1.2cm}|P{1.2cm}|P{1.2cm}|} \hline
Atoms ($\X$)      & K  & Cl    & H & O &  C  \\  \hline 
$\epsilon_\X$ (meV)  & 3.773    &  6.505   & 1.89 & 6.596   &3.036   \\ \hline
$r_\X$ (nm) &    0.176 &   0.227 &0.022  & 0.177 &    0.199    \\ \hline
\end{tabular}
\caption{\label{vdw} Lennard-Jones parameters for individual elements. The vdW potential between two atoms at a distance $d$, is calculated using the Lennard-Jones relation,  $V_\mathrm{LJ} = \epsilon_{m}\left[\left(\frac{r_{m}}{d}\right)^{12}-2\left(\frac{r_{m}}{d} \right)^6\right]$, where the parameter $r_{m}=r_1+r_2$ is the equilibrium distance and $\epsilon_{m}=\sqrt{\epsilon_1 \epsilon_2}$ is the well depth of the interaction.}
\end{table} 

%\vspace{\fill}

%\bibliography{reference}